\documentclass[twocolumn,english,prl]{revtex4-1}
\usepackage[T1]{fontenc}
\usepackage[latin9]{inputenc}
\usepackage{graphicx}
\usepackage{amsmath}
\usepackage[colorlinks = true,
linkcolor = blue,
urlcolor  = blue,
citecolor = blue,
anchorcolor = blue]{hyperref}
\usepackage[usenames, dvipsnames]{color}

\makeatletter
\usepackage{babel}
\usepackage{float}

\usepackage{tikz}
\usetikzlibrary{shapes.geometric, arrows}

\usepackage{tabularx}
\newcolumntype{C}{>{\centering\arraybackslash}X}

\newcommand{\angstrom}{${\buildrel _{\circ} \over {\mathrm{A}}}$}
\makeatother

\begin{document}
	
%
%
%
%
%

	\title{\textit{PyCrystalField}: Software for Calculation, Analysis, and Fitting of Crystal Electric Field Hamiltonians}
	
	\author{A. Scheie}
\email{scheieao@ornl.gov}
\address{Neutron Scattering Division, Oak Ridge National Laboratory, Oak Ridge, Tennessee 37831, USA}

\date{\today}

\begin{abstract}
	
We introduce \textit{PyCrystalField}, a Python software package for calculating single-ion crystal electric field (CEF) Hamiltonians. This software can calculate a CEF Hamiltonian \textit{ab initio} from a point charge model for any transition or rare earth ion in either the $J$ basis or the $LS$  basis, perform symmetry analysis to identify nonzero CEF parameters, calculate the energy spectrum and observables such as neutron spectrum and magnetization, and fit CEF Hamiltonians to any experimental data. The theory, implementation, and examples of its use are discussed.

\end{abstract}

\maketitle

\section{Introduction}

A material's electronic properties---which include magnetic, optical, and electric qualities---are governed by electron orbitals. Many effects determine an ion's precise orbital state, but one of the key effects is crystal electric field (CEF) interactions. These occur when Pauli exclusion and Coulomb repulsion from surrounding atoms shift the relative energies of valence electron orbitals. This can have dramatic effects on an ion's magnetism, causing single-ion anisotropies, high or low magnetic quantum numbers, or nonmagnetic singlets \cite{AbragamBleaney}.  To aid in setting up and carrying out such calculations, we introduce \textit{PyCrystalField}: a python library for calculating and fitting the CEF Hamiltonian of ions.

The theory behind CEF calculations was worked out over 50 years ago \cite{Stevens1952,Hutchings1964}, but  calculating the single-ion Hamiltonian remains a challenge for two reasons. First, the exact Hamiltonian cannot be calculated \textit{ab initio}. One can use an approximate "point charge model" which neglects higher-order effects \cite{EDVARDSSON1998,Mesot1998}, but a quantitatively accurate CEF Hamiltonian must be obtained by fitting to data. Second, the complexity of the CEF Hamiltonian depends upon the symmetry of the ion's environment, causing low-symmetry Hamiltonians to have many independent parameters. Because of this, determining a CEF Hamiltonian can be quite challenging.

Other software has been developed to do CEF calculations, such as \textit{CFcal} \cite{CFcal},  \textit{FOCUS} \cite{FOCUS}, \textit{Mantid} \cite{Mantid}, \textit{McPhase} \cite{McPhase}, \textit{SIMPRE} \cite{SIMPRE}, or \textit{SPECTRE} \cite{SPECTRE}. Useful as they are, these programs are limited in either the type of ions they consider---many point charge models are limited to rare earth ions in the $J$ basis---or the type of data they can fit. 
\textit{PyCrystalField} expands the flexibility of CEF software by allowing the user to fit any kind of data: the Python implementation gives the ability to define a custom $\chi^2$ function and fit any variable (including terms in point charge models). It also improves user-friendliness, by automatically building point charge models from .cif files (files which detail the crystal structure of a material \cite{Hall1991}) and fitting to data in just a few lines of code.
Finally, \textit{PyCrystalField} extends the range of point charge modeling to any ion, including to rare earth ions in the $LS$ basis.

\section{Theory}

The general CEF Hamiltonian can be written as:
\begin{equation}
\mathcal{H}_{CEF} =\sum_{n,m} B_{n}^{m}O_{n}^{m},
\end{equation}
where $O_{n}^{m}$ are the Stevens Operators \cite{Stevens1952,Hutchings1964} and $B_{n}^{m}$ are multiplicative factors called CEF parameters. $n$ is the operator degree, and is constrained by time-reversal symmetry to be even \cite{Newman_1971}. $m$ is the operators order, and $-n \geq m \geq n$. These parameters can either be fit to experimental data or calculated from a point charge model approximation as a starting point of the fit.

\subsection{Point Charge Model}

An approximate CEF Hamiltonian can be calculated by treating the surrounding ligands as point charges and calculating the CEF Hamiltonian from Coulombic repulsion. \textit{PyCrystalField} calculates $B_{n}^{m}$ using the method outlined by Hutchings \cite{Hutchings1964}, where
\begin{equation}
B_{n}^{m}=-\gamma_{nm}q C_{nm}\left\langle r^{n}\right\rangle \theta_{n}.
\label{eqA:CEFparams}
\end{equation}
Here, $\gamma_{nm}$ is a term calculated from the ligand environment expressed in terms of tesseral harmonics, $q$ is the charge of  the central ion (in units of $e$), $C_{nm}$ are normalization factors of the spherical harmonics, $\left\langle r^{n}\right\rangle$ is the expectation value of the radial wavefunction (taken from Edvarsson \cite{EDVARDSSON1998} for all rare-earth ions and Haverkort \cite{HaverkortPhdthesis} for $3d$ and $4d$ transition ions), and $\theta_{n}$ are constants
associated with electron orbitals of the magnetic ion. Equation \ref{eqA:CEFparams} is derived in Appendix \ref{PC_CEF_B}.

\subsection{Neutron Cross Section}

Neutron scattering is a common way to measure low-energy (meV range) CEF transitions from one state to another as it can disentangle phonon scattering from crystal field excitations. \textit{PyCrystalField} calculates the powder-averaged neutron cross section with the dipole approximation
\begin{multline}
\frac{d^2\sigma}{d\Omega d\omega} = N (\gamma r_0)^2 \frac{k'}{k}F^2(\mathbf{Q}) e^{-2W(\mathbf{Q})} \\
p_n|\langle \Gamma_m|\hat J_{\perp}|\Gamma_n \rangle|^2 \delta(\hbar \omega + E_n - E_m)
\label{eqA:NeutronCrossSec}
\end{multline}
\cite{Furrer2009neutron}, where the three elements of $N (\gamma r_0)^2$ are normalization factors, $k$ and $k'$ are the incoming and outgoing neutron wavevectors, $F(\mathbf{Q})$ is the electronic form factor, $e^{-2W(\mathbf{Q})}$ is the Debye Waller factor, $p_n$ is the Boltzmann weight, and $|\langle \Gamma_m|\hat J_{\perp}|\Gamma_n \rangle|^2  = 
\frac{2}{3} \sum_{\alpha} |\langle \Gamma_m|\hat J_{\alpha}|\Gamma_n \rangle|^2 $  is computed from the inner product of total angular momentum $J_{\alpha}$  ($\alpha = x,y,z$) with the CEF eigenstates $|\Gamma_n \rangle$.

In practice, the delta function $\delta(\hbar \omega + E_n - E_m)$ in eq. \ref{eqA:NeutronCrossSec} has a finite width due to the intrinsic energy resolution of the instrument and the finite lifetime of the excited states. By default, \textit{PyCrystalField} approximates the instrument resolution with a Gaussian profile and the finite lifetimes with a Lorentzian profile, making the effective profile a convolution of the two (\textit{PyCrystalField} uses a Voigt profile for the sake of computational efficiency). However, the user can specify a custom neutron-peak profile, also with an arbitrary peak-shape dependence on mode energy (This way the user can model asymmetric peaks, where the asymmetry and mode width depend upon energy \cite{IKEDA1985}). With either default or custom peak shapes, \textit{PyCrystalField} allows the user to specify an arbitrary resolution function giving the FWHM as a function of $\Delta E$. (A useful approximation for time-of-flight spectrometers can be found in  Windsor \cite{windsor1981pulsed}.)

\textit{PyCrystalField} can calculate 2D (intensity vs $\Delta E$) and 3D (intensity vs $\Delta E$ and $Q$) data sets, as shown in Fig. \ref{flo:NdMg_FittedData}. For 3D data sets, the ion and Debye-Waller factor must be specified in order to calculate $Q$-dependence. (At the time of writing, the 3D neutron data calculations are only available for rare-earth ions.)

\subsection{Magnetization and Susceptibility}

\sloppy It can be useful to compute susceptibility and magnetization from the CEF Hamiltonian. To do this, \textit{PyCrystalField} calculates magnetization non-perturbatively as $M_{\alpha} = g_J \langle J_{\alpha} \rangle$, where $\langle J_{\alpha} \rangle = \sum_i e^{\frac{-E_i}{k_B T}}\langle i \rvert J_{\alpha} \lvert i \rangle ~/ Z~$ and $|i\rangle$ are the eigenstates of the effective Hamiltonian ${\cal H} = {\cal H}_{CEF} + \mu_B g_J {\bf B}\cdot {\bf J} $, where ${\bf B}$ is magnetic field. Susceptibility $\chi_{\alpha, \beta} = \frac{\partial M_{\alpha}}{\partial B_{\beta}}$ is calculated via a numerical derivative of magnetization with respect to field.

The advantage of the non-perturbative approach is that it can be extended to large magnetic fields and high temperatures without sacrificing accuracy.

\subsection{Intermediate Coupling Scheme}

For rare earth ions, the crystal fields are much weaker than spin-orbit coupling and are generally treated as a perturbation to the spin-orbit Hamiltonian, operating on an effective spin $J$. However, for transition ions (and certain rare earths) where the spin-orbit Hamiltonian $\mathcal{H}_{SOC}=\lambda S \cdot L$ is of the same magnitude as $\mathcal{H}_{CEF}$, it is necessary to treat both CEF and spin-orbit coupling non-perturbatively in what is called the "intermediate coupling scheme" where $\mathcal{H}_{CEF}$ acts only on orbital angular momentum $L$ \cite{AbragamBleaney}.

\textit{PyCrystalField} can do all calculations in either the weak coupling scheme ($J$ basis) or the intermediate coupling scheme ($LS$ basis), although in the latter case the ion's spin-orbit coupling $\lambda$ must be provided by the user. The advantage of the intermediate scheme is that it can account for intermultiplet transitions, but the disadvantage is that the eigenkets are usually harder to interpret. 

In calculating the intermediate-coupling neutron spectrum, $|\langle \Gamma_m|\hat J_{\perp}|\Gamma_n \rangle|^2  = |\langle \Gamma_m|\hat L_{\perp} + \hat S_{\perp}|\Gamma_n \rangle|^2 $ and in calculating magnetization $M_{\alpha} = g_J \langle J_{\alpha} \rangle = \langle L_{\alpha} + g_e S_{\alpha} \rangle$. The formula for $f$-electron point charge constants $\theta_n$ have not been previously published, Users should note that \textit{PyCrystalField} neglects small interactions like the on-site electron Coulomb interaction. This may cause slight discrepancies between the calculated eigenstates from \textit{PyCrystalField} and other software, and it means that \textit{PyCrystalField} does not calculate Hund's coupling. (Accordingly, the user must specify the $S$ and $L$ values for transition ions.)

\subsection{$g$-tensor}

\textit{PyCrystalField} can calculate the Land\`{e} $g$-tensor from an intermediate coupling CEF Hamiltonian. 
The $g$ tensor is defined such that 
\begin{equation}
\mathcal{H}_{Zeeman} = \mu_B ({\bf B} \cdot {\bf L} + g_e  {\bf B} \cdot {\bf S}) = \mu_B {\bf B} \cdot g \cdot {\bf J},
\end{equation} 
where ${\bf B}$ is applied magnetic field, $\bf L$ is orbital angular momentum, $\bf S$ is spin angular momentum, and $\bf J$ is total angular momentum. With a doublet ground state $|\pm \rangle$, the Hamiltonian can be re-written using the Pauli spin matrices. Assuming a magnetic field along $z$ for simplicity
$$
\mathcal{H}_{eff} = \mu_B B_z \begin{bmatrix}
\langle + | L_z + g_e S_z | + \rangle & \langle + | L_z + g_e S_z | - \rangle \\
\langle - | L_z + g_e S_z | + \rangle & \langle - | L_z + g_e S_z | - \rangle \\
\end{bmatrix}
$$
in the $LS$ basis and
$$
\mathcal{H}_{eff} = \frac{1}{2} \mu_B B_z \begin{bmatrix}
g_{zz} & (g_{zx} - i g_{zy}) \\
(g_{zx} + i g_{zy}) & - g_{zz} 
\end{bmatrix}
$$
in the effective $J$ basis. Setting these two equations equal to each other and assuming $B_z$ is small gives the $g$ tensor values:
\begin{align}
\begin{split}
g_{zz} =& 2 \langle + | L_{z} + g_e S_{z} |+ \rangle,\\ g_{zx} + i g_{zy}=&2 \langle - | L_{z} + g_e S_{z} |+ \rangle,
\label{eq:gtensor}
\end{split}
\end{align}
and so on for $g_{yy}$, $g_{yx}$, etc. \textit{PyCrystalField} uses Eq. \ref{eq:gtensor} to compute the $g$ tensor from any intermediate coupling Hamiltonian.
For the weak coupling scheme used for rare-earth ions, the equation is the same but with $L_{\alpha} + g_e S_{\alpha}$ replaced by $J_{\alpha}$.

\section{Implementation}

\textit{PyCrystalField} is a collection of Python objects and functions which allow the user to build and fit CEF Hamiltonians with just a few lines of code. The workflow is summarized in Fig. \ref{flo:workflow}. The ability to write scripts greatly streamlines the analysis process. It is available for download at \url{https://github.com/asche1/PyCrystalField}. 
Please report bugs to scheieao@ornl.gov.

\tikzstyle{input} = [rectangle, rounded corners, text width=1.7cm, minimum height=1.1cm,text centered, draw=black, fill=green!30]
\tikzstyle{model} = [rectangle, rounded corners, text width=2.3cm, minimum height=1.1cm,text centered, draw=black, fill=blue!22]
\tikzstyle{fit} = [rectangle, rounded corners, text width=1.8cm, minimum height=1.1cm, text centered, draw=black, fill=orange!30]
\tikzstyle{output} = [rectangle, rounded corners, text width=3.5cm, minimum height=0.9cm,text centered, draw=black, fill=red!30]

\tikzstyle{arrow} = [thick,->,shorten >=0.1cm,shorten <=0.1cm]
\tikzstyle{curvedarrow} = [thick,->,shorten >=0.13cm,shorten <=0.13cm]

\begin{figure}
	\centering
	%
	
	\begin{tikzpicture}[node distance=1.9cm]
	\node (cif) [input] {.cif file {\color{gray} \texttt{\scriptsize importCIF}}};
	\node (PointCharge) [model, right of=cif, xshift=1.2cm] {Point charge model \\ {\color{gray} \texttt{\scriptsize Ligands}} \color{gray} \scriptsize class};
	\node (fitPC) [fit, right of=PointCharge, xshift=1.2cm] {fit point charge};
	\node (cef) [input, below of=cif] {CEF parameters};
	\node (Hamiltonian) [model, right of=cef, xshift=1.2cm] {CEF \\Hamiltonian \\ { \texttt{\color{gray} \scriptsize CFLevels}} \color{gray} \scriptsize class};
	\node (fitH) [fit, right of=Hamiltonian,  xshift=1.2cm] {fit CEF parameters};
	\node (observables) [output, below of=Hamiltonian] {Calculate Observables};
	
	\draw [arrow] (cif) -- (PointCharge);
	\draw [arrow] (cef) -- (Hamiltonian);
	\draw [arrow] (PointCharge) -- (Hamiltonian);
	\draw [arrow] (Hamiltonian) -- (observables);
	\draw [curvedarrow] (PointCharge) to [out=28,in=150] (fitPC);
	\draw [curvedarrow] (fitPC) to [out=210,in=-28] (PointCharge);
	\draw [curvedarrow] (Hamiltonian) to [out=28,in=150] (fitH);
	\draw [curvedarrow] (fitH) to [out=210,in=-28] (Hamiltonian);
	\end{tikzpicture}
	
	\caption{\textit{PyCrystalField} workflow. One can begin either with a .cif file or a list of CEF parameters, and then calculate a point charge model or a CEF Hamiltonian respectively. One can fit these models, or directly compute observables like neutron spectrum, susceptibility, etc.}
	
	\label{flo:workflow}
\end{figure}
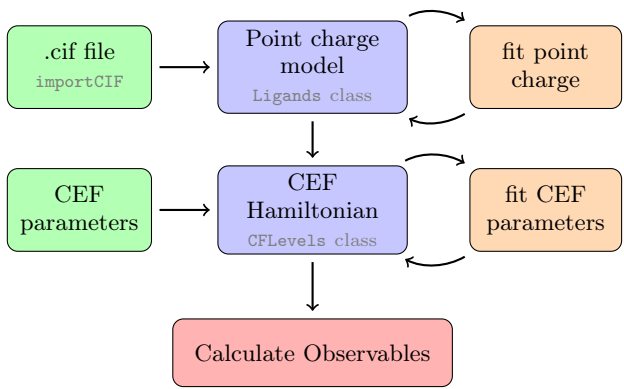

\subsection{Building CEF Hamiltonian}

The user may build a CEF Hamiltonian in two ways: (a) importing a crystal structure for a point charge model, or (b) specifying the CEF parameters $B_n^m$. 
For the former case, the user can import the structure from a .cif file or manually specify the locations of point charges around a central ion. \textit{PyCrystalField}'s importCIF function automatically analyzes the crystal symmetry to identify nonzero CEF parameters for the central ion, and it aligns the local coordinates so as to minimize the number of fitted CEF parameters (though the user can define custom axes as well). Specifically, \textit{PyCrystalField} places the $y$ axis normal to a mirror plane (if one exists) to eliminate the imaginary CEF operators and the $z$ axis along the highest-fold rotation axis if a rotation axis exists \cite{WALTER1984}. (Note that in low-symmetry point groups where one must choose between the $y$ axis alignment and the $z$ axis alignment, \textit{PyCrystalField} prioritizes having the $y$ axis normal to a mirror plane to avoid roundoff errors  with complex floating point calculations, unlike the formalism in Walter \cite{WALTER1984} which prioritizes having the $z$ axis along a rotation axis. This leads to slightly different sets of CEF parameters in very low-symmetry point groups, but the Hamiltonian is equivalent and the total number of CEF parameters is the same.)
\textit{PyCrystalField} can also import any range of neighboring ions to be included as point charges.

For magnetic rare-earth 3+ ions with orbital moments ($\rm Pr^{3+}$, $\rm Nd^{3+}$, $\rm Pm^{3+}$, $\rm Sm^{3+}$, $\rm Tb^{3+}$, $\rm Dy^{3+}$, $\rm Ho^{3+}$, $\rm Er^{3+}$, $\rm Tm^{3+}$, and $\rm Yb^{3+}$) the ground state $J$, $L$, $S$, and $\langle r^{n} \rangle$ values are automatically read from internal tables by \textit{PyCrystalField}. These internal tables also cover the common magnetic $3d$ and $4d$ ions (Cu$^{2+}$, Ni$^{2+}$, Ni$^{3+}$, Co$^{2+}$, Co$^{3+}$, Fe$^{2+}$, Fe$^{3+}$, Mn$^{2+}$, Mn$^{3+}$, Mn$^{4+}$, Cr$^{2+}$, Cr$^{3+}$, V$^{2+}$, V$^{3+}$, Ti$^{2+}$, Ti$^{3+}$, Nb$^{3+}$, Tc$^{4+}$, Ru$^{3+}$, Rh$^{3+}$, Pd$^{2+}$, and Pd$^{3+}$) and include spin-orbit coupling constants from Koseki \textit{et al.,} \cite{Koseki_2019} for calculations in the $LS$ basis. For other ions, $\langle r^{n} \rangle$ must be provided by the user, with the radial integral in units of Bohr radius $(a_0)^n$. 
Calculations in the $J$ basis are only available for rare-earth ions.

\subsection{Fitting}

\textit{PyCrystalField} allows complete freedom in fitting data. The user must provide a global $\chi^2$ function and \textit{PyCrystalField} will minimize $\chi^2$ by varying any user-specified variables. The user may fit any set of variables to the data: $B_n^m$, point charge location (demonstrated in Scheie \textit{et al} \cite{scheie2020crystal}; a similar method is discussed in Dun \textit{et al} \cite{dun2020effective}), effective charge of the point charges, width of neutron scattering peaks, etc. Any constraints may be imposed. The user may fit to any observed value: neutron data, magnetization, a list of eigenvalues, $g$-tensor anisotropy, or magnetic susceptibility. In this way, any variable may be fit to any kind of data.

The built-in fit functions are based on the Scipy minimize library, so a variety of methods are available. The \textit{PyCrystalField} fit function also makes it easy to add more variables to the set of fitted parameters, which is useful for sequentially fitting variables to complex data.

\section{Discussion}

\subsection{Examples}

In this section we present and discuss three examples to show the performance and capabilities of \textit{PyCrystalField}.

\subsubsection{Comparison to Manual Calculation: $\rm Yb^{3+}$ inside a cube of $\rm O^{2-}$ ions}

To confirm the accuracy of \textit{PyCrystalField}, we compare its results to a simple result calculated by hand. We computed the weak-coupling point charge CEF Hamiltonian for a $\rm Yb^{3+}$ ion surrounded by a cube of $\rm O^{2-}$ ions at a distance of $\sqrt{3}$ \angstrom$\>$ from the central ion. This system is easily calculable by hand following Hutchings (1964).

According to Hutchings' Eq. (6.11), a point charge cubic environment has a Hamiltonian of the form 
\begin{equation}
H = B_4^0 (O_4^0 + 5 O_4^4) +  B_6^0 (O_6^0 - 21 O_6^4),
\end{equation}
where for eight-fold coordination (i.e., a cube of ligands) $B_4^0 = \frac{7}{18}\frac{|e|q}{d^5}\beta_J \langle r^4 \rangle$ and $B_6^0 = -\frac{1}{9}\frac{|e|q}{d^7}\gamma_J \langle r^6 \rangle$ (Hutchings Table XIV), and 
for Yb$^{3+}$ 
$\beta_J = \frac{-2}{3 \cdot 5 \cdot 7 \cdot 11}$ and $\gamma_J = \frac{2^2}{3^3 \cdot 7 \cdot 11 \cdot 13}$ (Hutchings Table VI).
The resulting CEF parameters are given in Table \ref{flo:YbCube_CEF_params}. \textit{PyCrystalField}'s point charge calculation agrees with the result perfectly.

\begin{table}
	\centering
	\caption{Calculated CEF parameters for a $\rm Yb^{3+}$ ion inside a cube of $\rm O^{2-}$ ions, by hand via the method in Hutchings \cite{Hutchings1964}, and then by \textit{PyCrystalField}.}
	\begin{tabular}{c|cc}
		\hline \hline
		$B_n^m$ (meV) & By Hand & \textit{PyCrystalField} \\
		\hline 
		$ B_4^0$ & 0.1515035 & 0.1515035 \\
		$ B_4^4$ & 0.7575173 & 0.7575173 \\
		$ B_6^0$ & 0.001604 & 0.001604 \\
		$ B_6^4$ & -0.0336841 & -0.0336841 \\
		\hline \hline
	\end{tabular}
	\label{flo:YbCube_CEF_params}
\end{table}

\subsubsection{Fits to neutron data}

\begin{figure}
	\centering\includegraphics[width=0.47\textwidth]{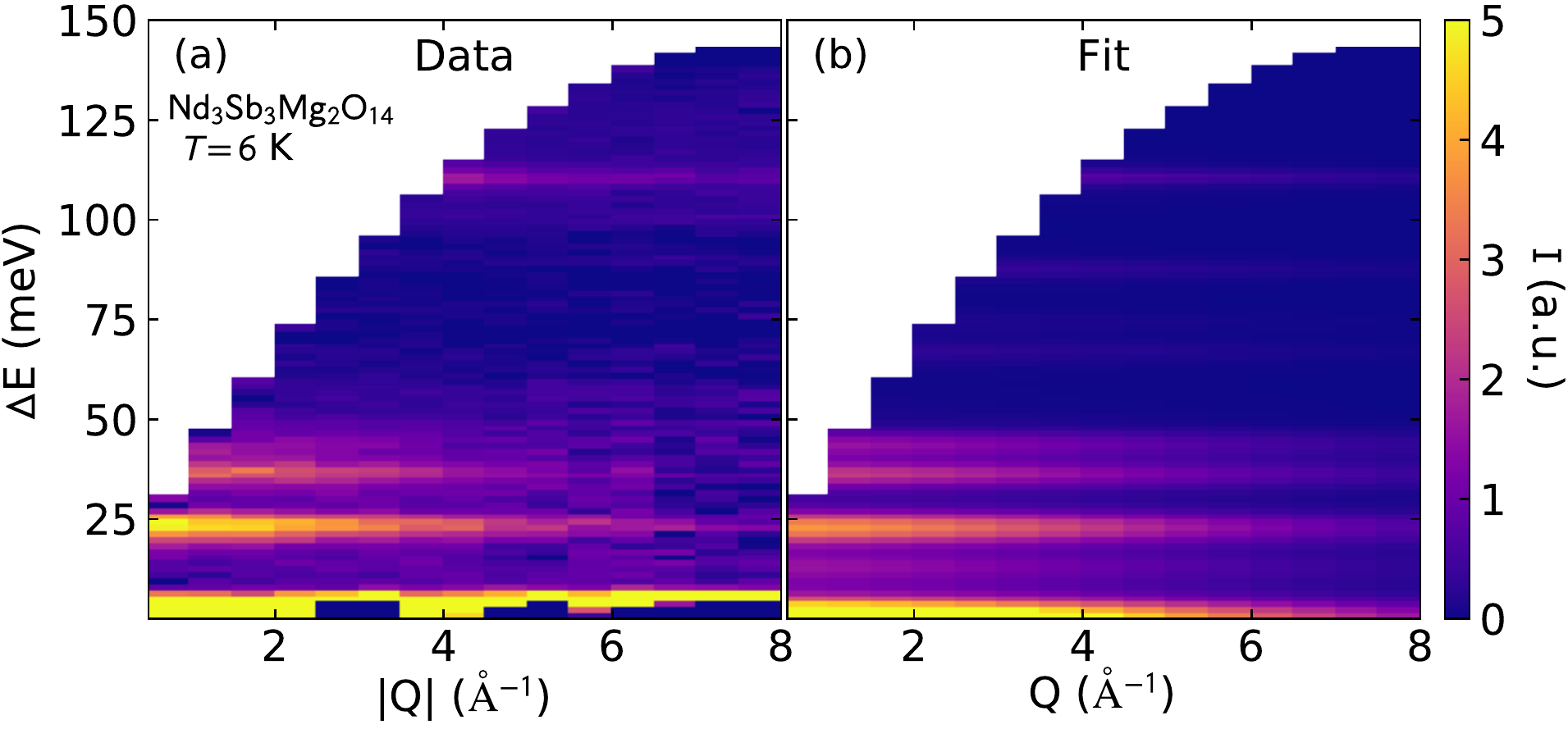}
	
	\caption{CEF fits to $\rm Nd_3Sb_3Mg_2O_{14}$ energy and momentum dependent neutron scattering data from Scheie \textit{et al} \cite{My_RE_KagCEF}. 
		(a) shows the data, and (b) shows the fit. The actual fit included three temperatures and three incident energies for a total of nine simultaneously fitted data sets.}
	
	\label{flo:NdMg_FittedData}
\end{figure}

\textit{PyCrystalField} was used to fit experimental neutron scattering data in Scheie \textit{et al} \cite{scheie2020crystal} and Scheie \textit{et al} \cite{My_RE_KagCEF}. 
It was used to fit the crystal field levels of a series of rare-earth "tripod" kagome materials ${\mathrm{Nd}}_{3}{\mathrm{Sb}}_{3}{\mathrm{Mg}}_{2}{\mathrm{O}}_{14}$ (shown in Fig. \ref{flo:NdMg_FittedData}), ${\mathrm{Nd}}_{3}{\mathrm{Sb}}_{3}{\mathrm{Zn}}_{2}{\mathrm{O}}_{14}$, and ${\mathrm{Pr}}_{3}{\mathrm{Sb}}_{3}{\mathrm{Mg}}_{2}{\mathrm{O}}_{14}$. These fits were done in two steps, fitting the effective charges of a point charge model, and then the CEF parameters directly to 2D $Q$ and $\Delta E$ dependent data at several energies and temperatures simultaneously \cite{My_RE_KagCEF}.

\begin{figure}
	\centering\includegraphics[width=0.46\textwidth]{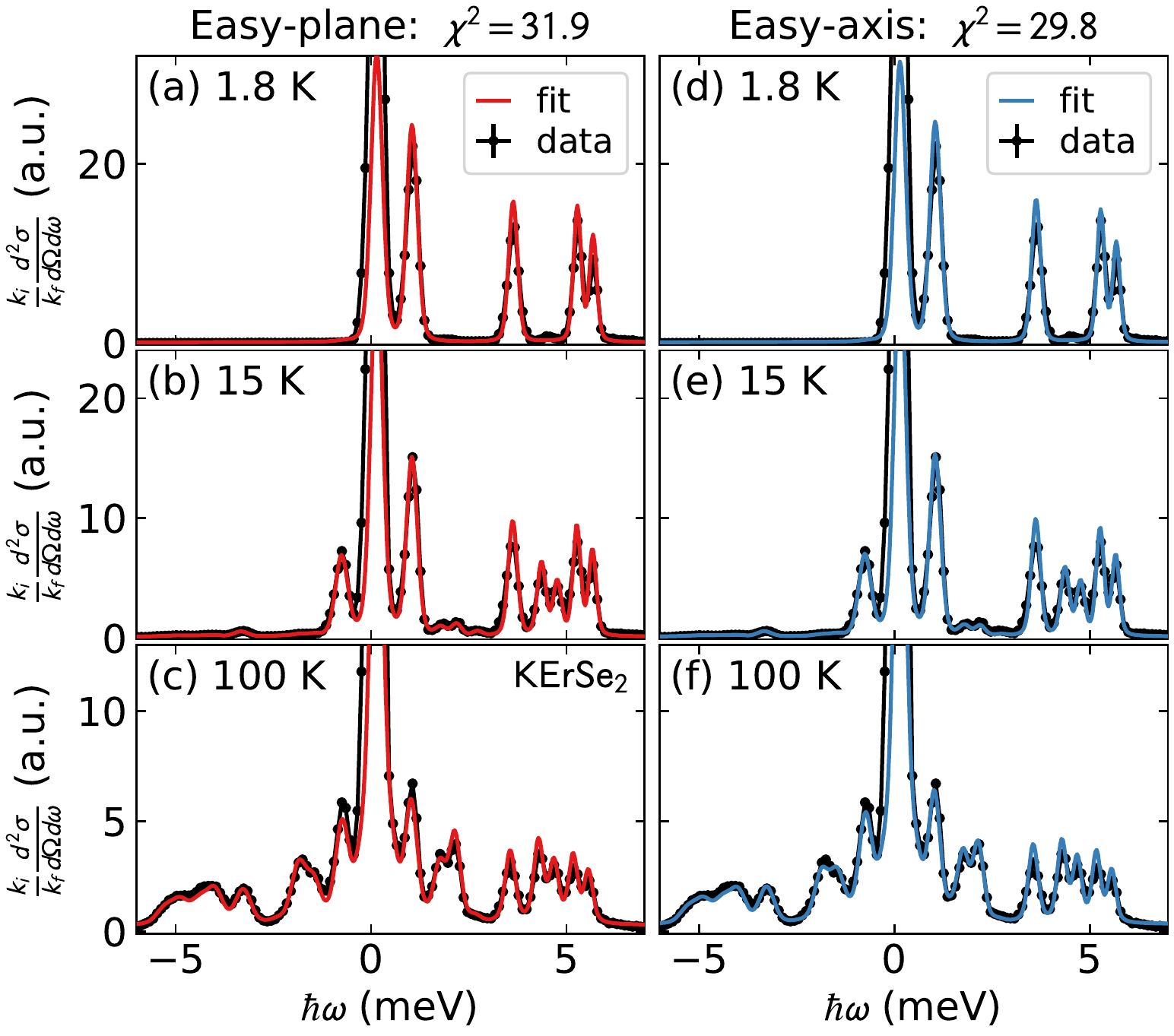}
	
	\caption{CEF fits to $\rm KErSe_2$ neutron scattering data from Scheie \textit{et al} (2020). 
		(a)-(c) show three different temperatures simultaneously fitted to an easy-plane model. (d)-(f) show the same data fitted to an easy-axis model. The fits are nearly identical, but bulk magnetization shows the easy-plane model to be correct.}
	
	\label{flo:KES_FittedData}
\end{figure}

\textit{PyCrystalField} was also used to fit the CEF Hamiltonian of Er triangular lattice delafossites $\rm KErSe_2$ and $\rm CsErSe_2$ \cite{scheie2020crystal}. An example fit is shown in Fig. \ref{flo:KES_FittedData}. This shows the ability of \textit{PyCrystalField} to fit multiple temperatures with overlapping peaks. This study also showcases the inherent limitations of CEF fits: two models emerge which fit the data beautifully but have opposite ground state anisotropies. In this case, bulk single-crystal magnetization was used to identify the correct model. This shows that CEF fits to neutron data can be underdetermined, even when fitting to many more peaks than parameters.

\subsubsection{Ab-initio Ni$^{2+}$ Energy Levels}

\textit{PyCrystalField} can also be used to calculate the ground state energy splitting of an ion in a particular ligand environment. If the ligand positions are well known, a point charge model is accurate enough to predict the general character of the ion's energy spectrum. This is similar to a comparison with a Tanabe-Sugano diagram \cite{AbragamBleaney}, but with more quantitative accuracy.

An example of this is the calculation of the electron orbital energies of $\rm Ni_2Mo_3O_8$, which has Ni$^{2+}$ ions surrounded by distorted octahedral and tetrahedral coordination of O$^{2-}$ ions \cite{JenHoneycomb}. \textit{PyCrystalField} allowed for \textit{ab initio} calculation of the energy spectrum of the valence orbitals in the intermediate coupling scheme. These calculations were used to qualitatively predict magnetic exchanges, single-ion anisotropies, and temperature dependence in $g$-factor within $\rm Ni_2Mo_3O_8$.

\subsection{Uses and Limitations}


Before using \textit{PyCrystalField}, users should be aware of the limitations and pitfalls of crystal field theory. First, the point charge model is imperfect because the ligand electrons generally have valence $p$-orbitals, which are anisotropic and cause the effective charge to differ from the idealized point charge model \cite{newman2007crystal}. Also, in itinerant systems, the conduction electrons can significantly alter CEF interactions, making the point-charge model even less reliable \cite{Birgeneau_1973}. Second, the success of fits involving many CEF parameters is highly dependent upon the starting values. There are various approaches to get around this, such as using the point charge model to generate starting parameters or using a Monte Carlo method to generate many sets of starting parameters. (In the latter case, a comparison to a point charge model is still helpful to ensure that the CEF parameters are physically sensible.)
Third, CEF fits can be underdetermined by neutron scattering data alone \cite{scheie2020crystal}: a fitted Hamiltonian should also be compared against independent measures of the anisotropy, such as bulk magnetization or electron spin resonance.
Fourth, the weak-coupling scheme and the intermediate-coupling scheme have different normalization factors (see Appendix \ref{IntermediateScheme}), requiring a re-calculation or re-fitting of the CEF parameters when translating between them.

\section{Conclusion}

\textit{PyCrystalField} is a general-purpose crystal field calculation package which allows the user to consider any type of data, any ion, and any ligand environment. It also allows the user to calculate observable quantities such as neutron scattering and bulk magnetization and susceptibility. It is accurate in reproducing simple point charge calculations, it can fit highly asymmetric Hamiltonians to complex data sets, and it can make qualitative ground state predictions from a point charge model.

\textit{PyCrystalField} simplifies difficult CEF calculations in a way that makes them practical for more of the scientific community. Understanding the single-ion properties in a material is often critical to understanding the behavior of the whole, and CEF calculations are a key part of this.

\section{Acknowledgments}

The author acknowledges helpful input from Collin Broholm, Youzhe Chen, Ovi Garlea, Tyrel McQueen, and Shan Wu.
This research at the Spallation Neutron Source was supported by the DOE Office of Science User Facilities Division. 
Initial stages of this work were supported by the Gordon and Betty Moore foundation under the EPIQS program GBMF4532. The author also acknowledges an anonymous referee who gave many helpful suggestions.

\appendix

\section{Calculation of Point charge CEF Parameters}\label{PC_CEF_B}

We begin with Hutchings \cite{Hutchings1964} Eq. 2.7:
$$
V(r,\theta,\phi)=\sum_{n=0}^{\infty}\sum_{\alpha}r^{n}\gamma_{n\alpha}Z_{n\alpha}(\theta,\phi),
$$
where $Z_{n\alpha}$ are tesseral harmonics, and 
\begin{equation}
\gamma_{n\alpha}=\sum_{j=1}^{k}\frac{4\pi}{(2n+1)}q_{j}\frac{Z_{n\alpha}(\theta_{j},\phi_{j})}{R_{j}^{n+1}},
\end{equation}
summing over $k$ ligands surrounding the central ion. 

Recognizing that (according to Hutchings' Eq. 5.3): 
$$V(x,y,z)=\sum_{mn}A_{n}^{m}\frac{1}{-|e|}f_{nm}^{c}(x,y,z)$$
and according to Hutchings' Eq. (5.5), the Hamiltonian can be written as
$$
\mathcal{H}=-|q|\sum_{i}V_{i}(x_{i},y_{i},z_{i})=\sum_{i}\sum_{mn}A_{n}^{m}f_{nm}^{c}(x,y,z),
$$
summing over electrons and where $q=Ze$. Alternatively, we can write the Hamiltonian
in terms of Stevens Operators:
$$
\mathcal{H}=-|Ze|\sum_{i}\sum_{n,m}r^{n}\gamma_{nm}Z_{nm}(\theta_{i},\phi_{i})
$$
$$
=\sum_{i}\sum_{n,m}A_{n}^{m}f_{nm}^{c}(x_{i},y_{i},z_{i})=\sum_{n,m}\underbrace{\left[A_{n}^{m}\left\langle r^{n}\right\rangle \theta_{n}\right]}_{B_{n}^{m}}O_{n}^{m}
$$
$$ \mathcal{H} =\sum_{n,m} B_{n}^{m}O_{n}^{m}, $$
where $\theta_{n}$ is a multiplicative factor which is dependent
on the ion ($\theta_{2}=\alpha_{J}$; $\theta_{4}=\beta_{J}$; $\theta_{6}=\gamma_{J}$;
see Table VI in Hutchings). Now, we solve the equations. 
We can look up $\left\langle r^{n}\right\rangle \theta_{n}$,
we just need to find $A_{n}^{m}$.

Because $A_{n}^{m}f_{nm}^{c}(x_{i},y_{i},z_{i})=-|e|r^{n}\gamma_{nm}Z_{nm}(\theta_{i},\phi_{i})$,
we should be able to find $A_{n}^{m}$. Now it turns out that, according
to Eq. 5.4 in Hutchings, $C_{nm}\times f_{nm}^{c}(x_{i},y_{i},z_{i})=r^{n}Z_{nm}^{c}(\theta_{i},\phi_{i})$,
where $C_{nm}$ is a multiplicative factor in front of the tesseral
harmonics. Therefore,
$$
A_{n}^{m}=-\gamma_{nm}^{c}|Ze|C_{nm}.
$$
A closed-form expression for the constants $C$ is very hard to derive,
but they are pre-calculated and can be found in "TessHarmConsts.txt".

In the end, the expression for crystal field parameters $B_{n}^{m}$ is
$$
B_{n}^{m}=A_{n}^{m}\left\langle r^{n}\right\rangle \theta_{n}
$$
\begin{equation}
B_{n}^{m}=-\gamma_{nm}|Ze|C_{nm}\left\langle r^{n}\right\rangle \theta_{n}.
\end{equation}
We know $|e|$, $Z$ (ionization of central ion), and the constants $C_{nm}$, $\theta_{n}$ are given in Hutchings, and $\left\langle r^{n}\right\rangle $ are found in Edvarsson \cite{EDVARDSSON1998}
 for rare earth ions.

\subsubsection*{Units}
PyCrystalField calculates the Hamiltonian in units of meV. $\gamma_{n\alpha}$ is in units of $\frac{e}{\textrm{\angstrom}^{n+1}}$,
$C_{nm}$ and $\theta_{n}$ are unitless, and $\left\langle r^{n}\right\rangle $
is in units of $(a_{0})^{n}$. This means that $B_{n}^{m}$ in the equation written above come out in units of $\frac{e^{2}}{\textrm{\angstrom}}$. 

To convert to meV, we first recognize that we have to re-write Hutchings
eq. (II.2) with the proper prefactor for Coulomb's law: $W=\sum_{i}\frac{1}{4\pi\epsilon_{0}}q_{i}V_{i}$. Thus,
our Hamiltonian becomes
$$
\mathcal{H}_{CEF}\,=\sum_{nm}B(exp)_{n}^{m}O_{n}^{m}=\sum_{nm}\frac{1}{4\pi\epsilon_{0}}B(calc)_{n}^{m}O_{n}^{m}.
$$
Now $\epsilon_{0}=\frac{e^{2}}{2\alpha hc}$, so
$$ 
\begin{aligned}
B(exp)_{n}^{m} & = \frac{1}{4\pi\epsilon_{0}}B(calc)_{n}^{m}=\frac{-1}{4\pi\epsilon_{0}}\gamma_{nm}|Ze|C_{nm}\left\langle r^{n}\right\rangle \theta_{n} \\
& =\frac{-2\alpha hc}{4\pi e^{2}}\left(\gamma_{nm}C_{nm}\left\langle r^{n}\right\rangle \theta_{n}\right)e^{2}Z\frac{a_{0}^{n}}{\textrm{\angstrom}^{n+1}}.
\end{aligned}
$$
Plugging in the values, we get the equation used by \textit{PyCrystalField}:
\begin{equation}
B_{n}^{m}=1.440\times10^{4}\left(0.5292\right)^{n}Z\left(\gamma_{nm}C_{nm}\left\langle r^{n}\right\rangle \theta_{n}\right){\rm meV}.
\end{equation}

\section{$\theta_n$ in the Intermediate Scheme} \label{IntermediateScheme}

When calculating crystal field levels from the point charge model, the $\theta_{n}$ constants are
listed in Stevens \cite{Stevens1952} 
for the ground states of all the rare earth
ions in the $J$ basis. But if we want to look at the state of the rare earth ion in
the $LS$ basis (intermediate coupling scheme), we must recalculate $\theta_{n}$ for each state.

General formulae for the first two constants, $\theta_{2}=\alpha$ and $\theta_{4}=\beta$,
are listed in Bleaney \& Stevens \cite{BleaneyStevens1953}. 
However, because they assume that such constants are only necessary
for $3d$ group ions, they do not list $\theta_{6}$. Therefore, we must derive it following the method of Stevens \cite{Stevens1952}. We take as an example the ${\rm Sm}^{3+}$ ion, which has five electrons in its $f$ orbital valence shell
and a ground state of ${\rm Sm}^{3+}$ has $S=5/2$, $L=5$, and $J=5/2$.

We know that 
\begin{equation}
\begin{aligned}
V_{6}^{0}&=\Sigma(231z^{6}-315z^{4}r^{2}+105z^{2}r^{4}-5r^{6}) \\
&=\gamma_{6,0}C_{6,0}\left\langle r^{6}\right\rangle \theta_{6}O_{6}^{0},
\end{aligned}\end{equation}
where the Stevens operator
\begin{equation}\begin{aligned}
O_{6}^{0}=&231L_{z}^{6}-(315X-735)L_{z}^{4} \\ & +(105X^{2}-525X+294)L_{z}^{2} \\
&-5X^{3}+40X^{2}-60X
\label{eq:StevensOp06}
\end{aligned}\end{equation}
and $X=L(L+1)$. Now to calculate $\theta_{6}$, we pick an eigenstate (in this case, $| L=5, S=5/2, m_l=5, m_s=-5/2 \rangle$) and calculate
the expectation value in terms of the Stevens Operators, and then
calculate it in terms of the individual electron wave functions. Then we set the two results equal to each other to find the multiplicative factor necessary to make the Stevens Operator result match the wave function integral.

\paragraph*{Stevens Operators:}
This is a straightforward calculation from eq. \ref{eq:StevensOp06}. Letting $G_{6}^{0}=\left\langle r^{6}\right\rangle \theta_{6}O_{6}^{0}$,
\begin{equation}\begin{aligned}
\langle L=5,\,m_{l}=5\rvert G_{6}^{0}\lvert L=5,\,m_{l}=5\rangle\, \\
=\,\theta_{6}\left\langle r^{6}\right\rangle 37800.
\label{eq:StevensOpCalc}
\end{aligned}\end{equation}

\paragraph*{Individual electron wave functions: }
To carry out this calculation we write $\lvert m_{l}=5,\,m_{s}=-\frac{5}{2}\rangle$ as a product
state of the individual electron wave functions in the valence shell,
which have $l=3$, $m=3,2,1,0,-1$ (adding up to $L=5$). Now we just recompute
$V_{6}^{0}$ in the basis of individual electrons: $V_{6}^{0}=\gamma O_{6}^{0}$
, where $m_{l}=m$ and $L=l$:
\begin{multline}
\langle L=5,\,m_{l}=5\rvert V_{6}^{0}\lvert L=5,\,m_{l}=5\rangle \\
=\,\{3,2,1,0,-1\}V_{6}^{0}\{3,2,1,0,-1\}\\
=\gamma(180-1080+2700-3600+2700)=900\gamma.
\label{eq:iewfCalc}
\end{multline}
We introduced $\gamma$ to account for the unknown scaling factor. (The similarity to $\gamma_{6,0}$ is unfortunate, because the $\gamma$s are unrelated. Nevertheless, we follow the notation of Stevens in this appendix.) So now we can relate eq. \ref{eq:StevensOpCalc}  to eq. \ref{eq:iewfCalc} so $\beta\left\langle r^{6}\right\rangle =\frac{900}{37800}\gamma=\frac{1}{42}\gamma$.
Now we find $\gamma$ by integrating the wave functions themselves. Let us pick the state $l=3$, $m=3$:
\begin{multline}
\langle l=3,m=3\rvert V_{6}^{0}\lvert l=3,m=3\rangle=180\gamma \\ =\int_{0}^{2\pi}\int_{0}^{\pi}Y_{3}^{3*}(231z^{6}-315z^{4}r^{2}+\\105z^{2}r^{4}-5r^{6})Y_{3}^{3}\sin\theta d\theta d\phi\\
=\frac{-80}{429}r^{6}.
\end{multline}
This means that $\gamma=-\frac{80}{180\cdot429}r^{6}=-\frac{4}{3861}r^{6}$, so that
\begin{equation}
\theta_{6}\left\langle r^{6}\right\rangle =-\frac{1}{42}\gamma=-\frac{4}{42\cdot3861}r^{6}
\end{equation}
and $\theta_6$ for ${\rm Sm}^{3+}$ is
\begin{equation}
\theta_{6}=-\frac{4}{42\cdot3861}.
\end{equation}

For each ion, we have to calculate $\langle L,\,m_{l}\rvert G_{6}^{0}\lvert L,\,m_{l}\rangle$
and then calculate $\{3,2...\}V_{6}^{0}\{3,2...\}$. Fortunately, the
result $\gamma_{6}=\frac{4}{3861}r^{6}$ holds for all rare earth
ions. Thus, we arrive at the general formula for $\theta_6$:

\begin{equation}
\theta_{6}=\frac{\{3,2...\}V_{6}^{0}\{3,2...\}}{\langle L,\,m_{l}\rvert O_{6}^{0}\lvert L,\,m_{l}\rangle}\,\frac{4}{3861}.
\end{equation}


%

\end{document}